\documentclass[aps,twocolumn]{revtex4}
\usepackage{epsfig}
\usepackage{graphicx}
\usepackage{amsmath}
\usepackage{booktabs}
\usepackage{makecell}
\usepackage{color}
\usepackage{multirow}

\begin{document}

\title{Hydrogenlike molecules composed of $D_1D_1$, $D_1D^*_2$ and $D^*_2D^*_2$}

\author{Hu-Hua He, Mao-Jun Yan, Chun-Sheng An, and Cheng-Rong Deng${\footnote{Contact author: crdeng@swu.edu.cn}}$}
\affiliation{School of Physical Science and Technology,
Southwest University, Chongqing 400715, China}

\begin{abstract}

We systematically explore the S-wave $D_1D_1$, $D_1D^*_2$ and $D^*_2D^*_2$ states with
various isospin-spin-orbit ($ISL$) configurations in the quark model. We propose nine
stable dimeson states with the $ISL$ configurations, $ISL=001$, $010$, $012$,
$100$, $102$, $110$, $112$, $120$, and $122$, against dissociation into their constituent
mesons. Those bound states are hydrogenlike molecular states, where the two subclusters are
moderately overlapped and the QCD covalent bond is formed due to the delocalization of light
quarks. The QCD covalent bond serves as the primary binding mechanism in the bound states
with $I=1$. However, the exchange of $\pi$ and $\sigma$-meson plays a pivotal role in the
bound states with $I=0$. The coupled-channel effect is essential in the formation of the
bound states with $ISL=001$, $010$, $012$, $100$, and $102$.

\end{abstract}

 \maketitle

\section{Introduction}

The history of studying multiquark
states can be traced back to the birth of the quark model in 1964~\cite{Gell-Mann:1964ewy}.
Exploring the natures of multiquark hadrons has been a critically important topic in hadronic
physics since Jaffe predicted the H-particle in 1977~\cite{Jaffe:1976yi}. These exotic states
may provide a wealth of low-energy strong interaction information that ordinary hadrons do
not~\cite{Jaffe:2004ph,Olsen:2017bmm}, offering valuable insights into the fundamental nature
of these interactions.

In 2003, the Belle Collaboration made a groundbreaking observation of the first narrow
hidden charm state very close to the $D\bar{D}^*$ threshold~\cite{Belle:2003nnu}, initially
named $X(3872)$ and now designated as $\chi_{c1}(3872)$~\cite{Gershon:2022xnn}. This
discovery marked a significant milestone in the study of heavy quarkonium states, opening
a gate for research into the nature of such kind of exotic states. In the following 20 years,
a large amount of hidden charm and hidden bottom hadrons, denoted as $XYZ$, $P_c$, and
$P_{cs}$ states, have been continuously observed in experiments~\cite{Chen:2016qju,Brambilla:2019esw}.

In 2021, the LHCb Collaboration reported the first observation of the doubly charmed
tetraquark state $T_{cc}(3875)^+$ in the $D^0D^0\pi^+$ invariant mass spectrum~\cite{LHCb:2021vvq,LHCb:2021auc}.
The deuteron-like molecular configuration $DD^*$ is manifest from its characteristic
size~\cite{LHCb:2021vvq,LHCb:2021auc}. Similar to the hidden charm family, it is
anticipated that there exists a diverse array of doubly charmed molecular tetraquarks
in hadron spectroscopy. Along these lines, the theorists proposed various
types of $T_{cc}(3875)^+$-like molecular states, including $T^{\prime}_{cc}$~\cite{Chen:2021vhg},
$T^-_{bb}$~\cite{Deng:2021gnb,Albaladejo:2021vln,Aoki:2023nzp},
$T^+_{ccs}$~\cite{Deng:2021gnb,Karliner:2021wju,Dai:2021vgf,Li:2023hpk,Tanaka:2024siw},
$T^0_{bc}$~\cite{Deng:2021gnb,Mathur:2021gqn,Alexandrou:2023cqg,Padmanath:2023rdu,Radhakrishnan:2024ihu},
$T^0_{bcs}$~\cite{Deng:2021gnb,Mathur:2021gqn,Liu:2023hrz}, $D_1D_1$~\cite{Wang:2021ajy,Chen:2024snh},
$D_1D_2^*$~\cite{Wang:2021ajy} and $D^*_2D_2^*$~\cite{Wang:2021ajy}. The LHCb experiment
as well as other experiments will provide more chances of observing the hadronic molecules.

One notable and crucial characteristic of the exotic states mentioned above is that
they are all positioned very near the thresholds of the hadron pairs to which they
can couple. Although the internal structure of these states remains uncertain and,
in some cases, controversial, the prevailing interpretation is that they are hadronic
molecular states, with binding energies ranging from several MeV to several tens
of MeV~\cite{Guo:2017jvc,Dong:2021bvy,Dong:2021juy,Chen:2022asf}. The binding mechanisms of
these states are complex and involve a range of interactions, including meson exchange,
short-range contact forces, and coupled-channel effects, etc. In molecular physics,
the covalent bond, formed through the overlap of atomic orbits and the sharing
of electrons between atoms, is essential in determining the structure and properties
of molecules. Similarly, is there a concept of a ``hadronic covalent bond" in the
doubly charmed tetraquark systems that resemble the bonding mechanism in the
hydrogen molecule?

In our previous study, we analyzed the underlying dynamics involved in the formation
of the loosely bound state $T_{cc}(3875)^+$ and identified the most promising candidates
for its partners, which are composed of two ground-state mesons within the quark model.
In this study, we extend our model investigation to S-wave dimeson states composed
of two excited charmed mesons, $D_1D_1$, $D_1D^*_2$, and $D^*_2D^*_2$, across various
isospin-spin-orbit configurations. Our objective is to explore potential bound states to
further enrich the doubly heavy tetraquark family. We will calculate their binding
energies, analyze their spatial configurations, and investigate the underlying binding
mechanisms. Through this work, we aim to provide valuable insights that could assist
in the experimental identification of doubly heavy tetraquark states in the future.

After the Introduction, the paper is organized as follows. In Sec. II, we outline
the key characteristics of the quark model. In Sec. III, we provide a brief overview
of the trial wave functions for mesons and dimeson states. In Sec. IV, we discuss
the properties of the dimeson ground states. Finally, in Sec. V, we conclude
with a summary.

\section{quark model}

The strong interactions are primarily described by quantum chromodynamics
(QCD) within the framework of the standard model of particle physics.
However, the {\it ab initio} calculation of hadron spectroscopy and hadron-hadron
interactions directly from QCD is quite challenging due to the complex nonperturbative
nature of the theory. This situation has led theoretical physicists to develop various
nonperturbative methods to elucidate the dynamics of strong interactions. Among
these methods, the quark models have been widely used for decades, offering several
advantages, including reduced computational complexity, a clear physical picture,
and strong predictive capabilities.

The constituent quark models have been established based on the assumption that hadrons
are colorless singlet, non-relativistic bound states of constituent quarks, characterized
by phenomenological effective masses and a variety of effective interactions. The origin
of the constituent quark mass can be traced back to the spontaneous breaking of
$SU(3)_L\otimes SU(3)_R$ chiral symmetry. As a result, the constituent quarks should
interact through the exchange of Goldstone bosons \cite{Manohar:1983md}. The chiral symmetry
is spontaneously broken in the light quark sector ($u$, $d$, and $s$) while being explicitly
broken in the heavy quark sector ($c$ and $b$). Consequently, the Goldstone bosons ($\pi$,
$K$, and $\eta$) exchange interactions occur only within the light quark sector. Additionally,
the scalar meson $\sigma$ exchange interaction is also involved in the model. The central
parts of the interactions can be resumed as follows~\cite{Vijande:2004he},
\begin{eqnarray}
\begin{split}
V_{ij}^{\rm obe}= & V^{\pi}_{ij} \sum_{k=1}^3\boldsymbol{F}_i^k
\boldsymbol{F}_j^k+V^{K}_{ij} \sum_{k=4}^7\boldsymbol{F}_i^k\boldsymbol{F}_j^k \\
+&V^{\eta}_{ij} (\boldsymbol{F}^8_i\boldsymbol{F}^8_j\cos \theta_P
-\sin \theta_P),\\
V^{\chi}_{ij}= &
\frac{g^2_{ch}}{4\pi}\frac{m^3_{\chi}}{12m_im_j}
\frac{\Lambda^{2}_{\chi}}{\Lambda^{2}_{\chi}-m_{\chi}^2}
\boldsymbol{\sigma}_{i}\cdot
\boldsymbol{\sigma}_{j}  \\
\times &\left( Y(m_\chi r_{ij})-
\frac{\Lambda^{3}_{\chi}}{m_{\chi}^3}Y(\Lambda_{\chi} r_{ij})
\right),~Y(x)=\frac{e^{-x}}{x} \\
V^{\sigma}_{ij}= &-\frac{g^2_{ch}}{4\pi}
\frac{\Lambda^{2}_{\sigma}m_{\sigma}}{\Lambda^{2}_{\sigma}-m_{\sigma}^2}
\left( Y(m_\sigma r_{ij})-
\frac{\Lambda_{\sigma}}{m_{\sigma}}Y(\Lambda_{\sigma}r_{ij})
\right).  \\
\end{split}
\end{eqnarray}

Besides the chiral symmetry breaking, it is expected that the dynamics of the model
are governed by QCD. The perturbative effect is primarily represented by the well-known
one-gluon exchange (OGE) interaction. From the nonrelativistic reduction of the OGE
diagram in QCD for point-like quarks, we obtain the following expression,
\begin{eqnarray}
V_{ij}^{\rm oge}={\frac{\alpha_{s}}{4}}\boldsymbol{\lambda}^c_{i}
\cdot\boldsymbol{\lambda}_{j}^c\left({\frac{1}{r_{ij}}}-
{\frac{2\pi\delta(\mathbf{r}_{ij})\boldsymbol{\sigma}_{i}\cdot
\boldsymbol{\sigma}_{j}}{3m_im_j}}\right),
\end{eqnarray}
where $\boldsymbol{\lambda}^c_{i}$ and $\boldsymbol{\sigma}_{i}$ represent
the color $SU(3)$ Gell-Mann matrices and the spin $SU(2)$ Pauli matrices,
respectively. Here, $r_{ij}$ denotes the distance between quarks $i$ and $j$,
and $m_i$ is the mass of the $i$-th quark.

Finally, any model that aims to mimic QCD must incorporate the nonperturbative
effect of color confinement. We adopt a phenomenological color screening
confinement potential to address this requirement,
\begin{eqnarray}
V^{\rm con}_{ij}&=&-a_c\boldsymbol{\lambda}^c_i\cdot\boldsymbol{\lambda}^c_jf(r_{ij})\nonumber\\
\noalign{\smallskip} f(r_{ij})&=&\left \{
\begin{array}{cccccc}
r^2_{ij}~~~~~~~~~\mbox{if $i$, $j$ occur in the same meson}, \nonumber\\
\noalign{\smallskip} \frac{1-e^{-\mu_c r^2_{ij}}}{\mu_c} ~\mbox{if
$i$, $j$ occur in different mesons.}\nonumber
\end{array}
\right.
\end{eqnarray}

This type of hybrid confinement potential arises from the quark delocalization
and color screening model~\cite{Wang:1992wi,Wang:1995bg}, which can effectively
describe the nuclear intermediate-range attraction. It also reproduces nucleon-nucleon
scattering data and the properties of the deuteron~\cite{Wang:1992wi,Wang:1995bg}.
Recently, this model has successfully described the properties of the $T^+_{cc}$
state observed by the LHCb Collaboration~\cite{Deng:2021gnb}.

\section{wave functions}

The wave function of a colorless charmed meson with isospin $I$ and angular momentum $J$ can
be expressed as the direct product of its constituent parts: the color part $\chi_c$, the
isospin part $\eta_{i}$, the spin part $\psi_s$, and the spatial part $\phi^G_{lm}(\mathbf{r})$.
This can be written mathematically as
\begin{eqnarray}
\Phi^{D}_{IJ}=\chi_c\otimes\eta_{i}\otimes\psi_s\otimes\phi^G_{lm}(\mathbf{r}),
\end{eqnarray}
where $D$ stands for all possible charmed mesons, encompassing both ground states and
excited states. $\mathbf{r}$ is the relative coordinate between the quarks $c$ and $\bar{q}$,
where $\bar{q}$ represent either $\bar{u}$ or $\bar{d}$ in the context of the $D$ meson.

To obtain a reliable solution to the few-body problem, a high-precision
numerical method is essential. The Gaussian Expansion Method (GEM) has
proven to be a highly effective tool for solving few-body
problems~\cite{Hiyama:2003cu,Hiyama:2018ivm}; thus, in this study, we
employ the GEM to investigate the doubly heavy tetraquark system.
According to the GEM, the relative motion wave function can be written as
a superposition of a set of Gaussian functions with specified angular momentum,
\begin{eqnarray}
\phi^G_{lm}(\mathbf{r})=\sum_{n=1}^{n_{max}}c_{n}N_{nl}r^{l}e^{-\nu_{n}r^2}
Y_{lm}(\hat{\mathbf{r}}),\nonumber
\end{eqnarray}
More details about the GEM can be found in Refs.~\cite{Hiyama:2003cu,Hiyama:2018ivm}.

The doubly charmed molecular states, denoted as $T_{cc}$, can be established
by the colorless mesons $D_1$ (composed of $c_1\bar{q}_1$) and $D_2$
(composed of $c_2\bar{q}_2$), where the indices just denote the specific particles
rather than their angular momentum. The wave function of the states with defined isospin $I$
and total angular momentum $J$ can be expressed as
\begin{eqnarray}
\begin{aligned}
\Psi^{T_{cc}}_{IJ}=\sum_{\xi}
c_{\xi}\mathcal{A}_{12}\left\{\left[\Phi^{D_1}_{I_1J_1}
\Phi^{D_2}_{I_2J_2}\right]_{IJ}\phi_{lm}^G(\boldsymbol{\rho})\right\}.
\end{aligned}
\end{eqnarray}
The term $\phi_{lm}^G(\boldsymbol{\rho})$
represents the wave function for the relative motion between the two mesons in the center-of-mass
frame. The Jaccobi coordinate $\boldsymbol{\rho}$ can be explicitly expressed as
\begin{eqnarray}
\begin{aligned}
&\boldsymbol{\rho}=\frac{m_c\mathbf{r}_{c_1}+m_q\mathbf{r}_{\bar{q}_1}}{m_c+m_{\bar{q}}}
-\frac{m_c\mathbf{r}_{c_2}+m_q\mathbf{r}_{\bar{q}_2}}{m_c+m_{\bar{q}}}.
\end{aligned}
\end{eqnarray}
Similarly, this wave function can be expanded as a superposition of a set of Gaussian functions.
In this work, we primarily focus on the ground state of our systems, as it is more likely to
form a bound state~\cite{Guo:2017jvc}.

The square brackets indicate the Clebsch-Gordan couplings of angular momentum and isospin. The
operator $\mathcal{A}_{12}$ serves as an antisymmetrization operator that acts on the identical
quarks $c_1$ and $c_2$, as well as on the identical anti-quarks $\bar{q}_1$ and $\bar{q}_2$.
\begin{eqnarray}
\begin{aligned}
&\mathcal{A}_{12}=(1-P_{c_1c_2})(1-P_{\bar{q}_1\bar{q}_2}),
\end{aligned}
\end{eqnarray}
where $P$ is the permutation operator acting on the identical particles. The summation index
$\xi$ encompasses all possible isospin-spin intermediate configurations $\{I_1, I_2, J_1, J_2\}$
that can be coupled to yield the total isospin $I$ and angular momentum $J$ of the state $T_{cc}$.
From a technical standpoint, the coefficients $c_{\xi}$ can be determined by solving the eigen
problem. Physically, they are dictated by the dynamics of the model and the inherent properties
of the dimeson states. To satisfy Bose-Einstein statistics, a restriction must be imposed on
the quantum numbers $\{I_1, I_2, J_1, J_2\}$ when the mesons $D_1$ and $D_2$ are identical
bosons. Their quantum numbers must satisfy the relation $J_1+J_2-J+I_1+I_2-I=\text{even}$.

\section{Numerical Results and Analysis}

\subsection{Methodology}

The first step in studying tetraquark systems is to incorporate ordinary mesons within the
quark model in order to determine the model parameters. To achieve this, we applied the MINUIT
program for fitting the mass spectrum of ground-state mesons~\cite{Deng:2014gqa}. This program
is widely used for fitting models to data, aiming to minimize an objective function and derive
the best-fit parameter values along with their associated uncertainties~\cite{James:1975dr}.

We minimize the mean square error
\begin{eqnarray}
\Delta=\sum_{i=1}^N\frac{w_i(M_i-m_i)^2}{N}\nonumber
\end{eqnarray}
to fit the mass spectrum and to determine the adjustable parameters and their errors
in the MINUIT program~\cite{Deng:2014gqa}. $N$ is the total number of mesons.
$M_i$ is the experimental mass of the ith meson and $m_i$ is its predicted mass in
the model. $w_i$ is its corresponding weight for fitting mass spectrum better.
For heavier mesons, the weights are set to 1, while for lighter mesons, particularly
the $\pi$ and $K$ mesons, the weights are set greater than 1, such as 2 or 3.
This objective function quantifies the goodness of fit between the model and the
observed data. When the model closely matches the data, minimizing the function
provides accurate estimates of the model parameters. A better fit results in smaller
uncertainties for these parameters.

The values of the model parameters and their uncertainties for the ground state
mesons are provided in Table II of Ref~\cite{Deng:2014gqa}. These surprisingly
small uncertainties relative to the model parameters arise from the synergy of an
efficient optimization process and the application of statistical methods
(the covariance matrix) in the MINUIT program to quantify these uncertainties.
Based on the model parameters obtained from fitting the ground-state meson spectrum~\cite{Deng:2014gqa},
we present the masses of the nonstrange charmed and bottomed excited mesons in Table~\ref{mesons}.
$D_0$ is absent because it has a very large width, which prevents it from being
a good building block of hadronic molecules. It is worth noting that the mass
difference between the experimental data and the model predictions for the excited
states is on the order of several tens of MeV. However, this discrepancy does not
significantly impact the binding energy of the dimeson states, due to the specific
definition of reduction used in the calculation, see Eq.~(\ref{eq:be}).

\begin{table}[h]
\caption{Nonstrange charmed and bottomed excited mesons, mass unit in MeV and
$\langle r^2\rangle^{\frac{1}{2}}$ unit in fm.}
\tabcolsep=0.25cm
\label{mesons}
\begin{tabular}{cccccccccccccccccc}
\toprule[0.8pt] \noalign{\smallskip}
State&$IJ^{P}$&Theory&Particle data group&$\langle \boldsymbol{r}^2\rangle^{\frac{1}{2}}$\\
\toprule[0.8pt] \noalign{\smallskip}
$D_1$&$\frac{1}{2}1^+$&2361$\pm$4&$2422.1\pm0.6$
&1.16\\
\noalign{\smallskip}
$D_2^*$&$\frac{1}{2}2^+$&2368$\pm$4&$2461.1\pm0.8$&1.17\\
\noalign{\smallskip}
$D_3^*$&$\frac{1}{2}3^-$&2677$\pm$4&$2763.1\pm3.2$&1.45\\
\noalign{\smallskip}
$B_1$&$\frac{1}{2}1^+$&5666$\pm$4&$5726.0^{+2.5}_{-2.7}$&1.13\\
\noalign{\smallskip}
$B_2^*$&$\frac{1}{2}2^+$&5668$\pm$4&$5737\pm0.7$&1.14\\
\noalign{\smallskip}\toprule[0.8pt]
\end{tabular}
\end{table}

The model Hamiltonian can be regarded as a function of model parameters, $H=H(x_1,...,x_8)$.
The variance of the model hamiltonian resulting from the parameter uncertainties $\delta x_i$
can be written as
\begin{eqnarray}
\delta H=\sum_{i=1}^{8}{\frac{\partial{H(x_1,...,x_8)}}{\partial{x_i}}}\delta x_i,
\end{eqnarray}
where $x_i$ and $\delta x_i$ represent the $i$-th adjustable parameter and it's error,
respectively, which are listed in Table II of Ref.~\cite{Deng:2014gqa}. The energy uncertainty
introduced by the parameter uncertainties can be calculated using the formula,
\begin{eqnarray}
\delta E=\langle\Phi_{IJ}\left|\delta H\right|\Phi_{IJ}\rangle,
\end{eqnarray}
where $\Phi_{IJ}$ is the eigen vector, i.e., $H\left|\Phi_{IJ}\rangle=E\right|\Phi_{IJ}\rangle$,
obtained by solving Shr\"{o}dinger equation. The energy uncertainties of ground state
mesons are presented in Table III of Ref.~\cite{Deng:2014gqa}. Those of nonstrange charmed
and bottomed excited mesons are listed in Table I, which are approximately 4 MeV.

Utilizing the quark model, we then investigate the state $T_{cc}(3875)^+$ observed by the LHCb
Collaboration. Our analysis indicates that the model describes the state $T_{cc}(3875)^+$
as a loosely bound, deuteron-like state within the quark framework, a description
that aligns remarkably well with the experimental data~\cite{Deng:2021gnb}.

Next, we turn our attention to the investigation of the dimeson states $T_{cc}$, composed
of two excited charmed states within the quark model. Generally, orbitally excited bound
states are more difficult to form due to the repulsive centrifugal potential. Therefore,
this study will focus on the ground states $T_{cc}$ with various isospin-spin-orbit
configurations, where two excited charmed mesons are in the relative $S$-wave. To obtain
the eigenvalues and eigenvectors of the dimeson states $T_{cc}$, we solve the four-body
Schr\"{o}dinger equation
\begin{eqnarray}
(H_4-E_4)\Psi^{T_{cc}}_{IJ}=0
\end{eqnarray}
for the bound-state problem with the Rayleigh-Ritz variational principle.

The binding energy $\Delta E_4$ of the states $T_{cc}$ is defined as
\begin{eqnarray}
\Delta E_4=E_4-\lim_{\rho \rightarrow \infty}E_4(\rho).\label{eq:be}
\end{eqnarray}
This expression is applied to determine whether the dimeson states $T_{cc}$ are stable against
the strong interactions. Here, $E_4(\infty)$ represents the theoretical threshold, i.e.,
$E(\infty)=E_{D_1}+E_{D_2}$, corresponding to the energy of two completely separated
mesons $D_1$ and $D_2$ that can couple to the same quantum numbers as the dimeson states
$T_{cc}$. If $\Delta E_4\geq0$, the dimeson state $T_{cc}$ is unbound and can decay into two
constituent mesons via strong interactions. However, if $\Delta E_4<0$, the strong decay
into two constituent mesons $D_1$ and $D_2$ is forbidden, and the decay can only occur via
weak or electromagnetic interactions. Their uncertainty resulting from the parameter
uncertainties can be written as
\begin{eqnarray}
\delta(\Delta E_4)=\delta E_4-\delta E_{D_1}-\delta E_{D_2}.
\end{eqnarray}

We present the binding energies of the dimeson states with various isospin-spin-orbit
configurations and their uncertainties in Table II. Whether in the dimeson $T_{cc}$ or
at its threshold, the uncertainties resulting from each parameter exhibit the same sign.
In the binding energy, the uncertainties resulting from each parameter can effectively
cancel each other out, resulting in a very small overall uncertainty. As a result, this
subtraction of binding energy significantly reduces the influence of uncertainties in
the model parameters and meson spectra on the calculated binding energies.

In the present calculation, we consider only the spin-spin and orbit-orbit
couplings, neglecting the spin-orbit coupling. Based on prior experience, the mass splitting
caused by the spin-orbit interaction is deemed negligible~\cite{Cleven:2015era,Deng:2019dbg}.
A more comprehensive investigation into the effects of spin-orbit coupling is left
for future studies.

\begin{figure*} [ht]
\centering
\resizebox{1.01\textwidth}{!}{\includegraphics{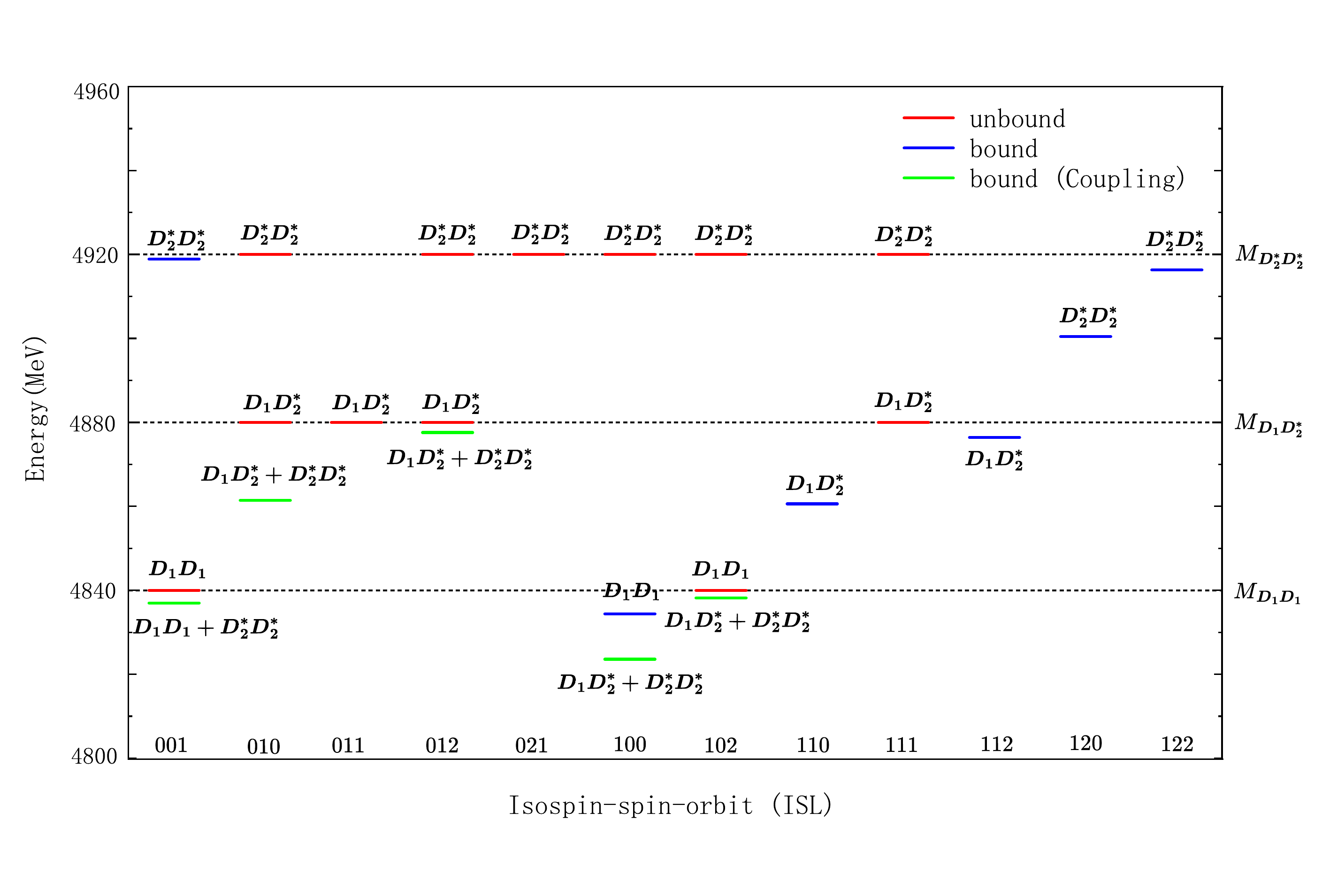}}
\caption{The energy spectrum of the dimeson states $D_1D_1$, $D_1D^*_2$, and $D^*_2D^*_2$
with various isospin-spin-orbit configurations. The thresholds $M_{D_1D_1}$, $M_{D_1D^*_2}$, and $M_{D^*_2D^*_2}$
are marked by three dotted horizontal lines. The red and blue line represents the single channel
while the green line stands for the result for the mixing of two channels with the same $ISL$.}
\label{spectrum}
\end{figure*}

To provide a clearer and more intuitive representation of
the energy spectrum, the binding energies of these dimeson states, relative to their
constituent mesons, are displayed in Fig.~\ref{spectrum}. The thresholds for $D_1D_1$,
$D_1D_2^*$, and $D_2^*D_2^*$ are indicated by dotted horizontal lines. This visualization
offers a better understanding of the relative binding energies and the positioning of each
state in relation to the respective thresholds.

To clarify the binding mechanism of the bound states, we analyze
the contributions from various interactions $\Delta\langle V^{\chi}\rangle$ and kinetic energy
$\Delta T$ to the binding energy $\Delta E_4$ using its eigenvector,
\begin{eqnarray}
\begin{aligned}
\Delta\langle V^{\chi}\rangle&=\langle\Psi^{T_{cc}}_{IJ}|V^{\chi}|\Psi^{T_{cc}}_{IJ}\rangle
-\langle\Phi^{D_1}_{I_1J_1}|V^{\chi}|\Phi^{D_1}_{I_1J_1}\rangle\\
&-\langle\Phi^{D_2}_{I_2J_2}|V^{\chi}|\Phi^{D_2}_{I_2J_2}\rangle,
\end{aligned}
\end{eqnarray}
where $\chi$ represents all types of interactions in the quark model. In order to reveal the
spatial configuration, we compute the size of the mesons and the distances between two mesons
in the dimeson state $T_{cc}$. The numerical results from these calculations are presented
in Table~\ref{tcc}.

\begin{table*}
\caption{Binding energy of the dimeson states relative to their constituents, its uncertainty and
the contribution from various interactions and kinetic energy, unit in MeV. $\Delta V^{\sigma}$,
$\Delta V^{\pi}$, $\Delta V^{\eta}$, $\Delta V^{\rm con}$, $\Delta V^{\rm cm}$, $\Delta V^{\rm coul}$,
and $\Delta T$ represent $\sigma$-, $\pi$-, and $\eta$-meson term, confinement term, chromomagnetic
term, Coulomb term, and kinetic energy, respectively. $I$, $S$ and $L$ represent total isospin, spin,
orbital angular momentum, respectively. $\langle\boldsymbol{r}^2\rangle^{\frac{1}{2}}$
is the size of mesons and $\langle\boldsymbol{\rho}^2\rangle^{\frac{1}{2}}$ is the distance between
two mesons, unit in fm.} \label{tcc}
\tabcolsep=0.24cm
\begin{tabular}{ccccccccccccccccccccccc}
\toprule[0.8pt] \noalign{\smallskip}
Constituent&$I,S\oplus L$&$\Delta E_4+\delta \Delta E_4$&$\langle\boldsymbol{r}^2\rangle^{\frac{1}{2}}$&$\langle\boldsymbol{\rho}^2\rangle^{\frac{1}{2}}$&$\Delta V^{\sigma}$
&$\Delta V^{\pi}$&$\Delta V^{\eta}$&$\Delta V^{\rm con}$&$\Delta V^{\rm cm}$&$\Delta V^{\rm coul}$&$\Delta T$\\
\noalign{\smallskip} \toprule[0.8pt] \noalign{\smallskip}
                             &$0,0\oplus 1$&Unbound\\
\noalign{\smallskip}

$D_1D_1$                     &$1,0\oplus 0$&{$-5.6\pm0.1$} &1.15&1.89&$-2.8$&$-0.3$&$-0.1$&$-2.8$&$-2.0$&$3.3$&$-0.9$ \\
\noalign{\smallskip}
                             &$1,0\oplus 2$&Unbound\\
\noalign{\smallskip}
\noalign{\smallskip}
                             &$0,1\oplus 0$&Unbound\\
\noalign{\smallskip}
                             &$0,1\oplus 1$&Unbound\\
\noalign{\smallskip}
\multirow{2}{*}{$D_1D_2^*$}  &$0,1\oplus 2$&Unbound\\
\noalign{\smallskip}
                             &$1,1\oplus 0$&{$-19.4\pm0.2$}&1.13&1.39&$-3.8$&$-1.3$&$-0.3$&$-8.7$&$-1.7$&$8.2$&$-11.8$\\
\noalign{\smallskip}
                             &$1,1\oplus 1$&Unbound \\
\noalign{\smallskip}
                             &$1,1\oplus 2$&{$-3.6\pm0.1$} &1.15&1.99&$-3.0$&$-1.0$&$-0.2$&$-1.1$&$-1.4$&$9.3$&$-6.3$ \\
\noalign{\smallskip}

\noalign{\smallskip}
\multirow{11}{*}{$D_2^*D_2^*$} &$0,0\oplus 1$&{$-1.1\pm0.1$}&1.17&2.78&$-1.8$&$-1.8$&$-0.0$&$-1.5$&$-2.1$&$0.3$&$5.8$ \\
\noalign{\smallskip}
                             &$0,1\oplus 0$&Unbound\\
\noalign{\smallskip}
                             &$0,1\oplus 2$&Unbound\\
\noalign{\smallskip}
                             &$0,2\oplus 1$&Unbound\\
\noalign{\smallskip}
                             &$1,0\oplus 0$&Unbound\\
\noalign{\smallskip}
                             &$1,0\oplus 2$&Unbound\\
\noalign{\smallskip}
                             &$1,2\oplus 0$&{$-19.6\pm0.2$}&1.14&1.40&$-3.8$&$-1.3$&$-0.3$&$-8.9$&$-1.9$&$8.2$&$-11.7$ \\
\noalign{\smallskip}
                             &$1,1\oplus 1$&Unbound\\
\noalign{\smallskip}
                             &$1,2\oplus 2$&{$-3.7\pm0.1$} &1.16&1.96&$-3.0$&$-1.0$&$-0.2$&$-1.2$&$-1.5$&$9.4$&$-6.3$ \\
\noalign{\smallskip}
\noalign{\smallskip} \toprule[0.8pt]
\end{tabular}
\caption{Channel coupled effects for the bound states with the same $ISL$ and sole bound channel, see the caption for Table~\ref{tcc}.} \label{coupling}
\tabcolsep=0.186cm
\begin{tabular}{ccccccccccccccccccccccc}
\toprule[0.8pt] \noalign{\smallskip}
Constituent&$I,S\oplus L$&Ratio&$\Delta E_4+\delta \Delta E_4$&$\langle\boldsymbol{r}^2\rangle^{\frac{1}{2}}$&$\langle\boldsymbol{\rho}^2\rangle^{\frac{1}{2}}$
&$\Delta V^{\sigma}$&$\Delta V^{\pi}$&$\Delta V^{\eta}$&$\Delta V^{\rm con}$&$\Delta V^{\rm cm}$&$\Delta V^{\rm coul}$&$\Delta T$\\
\noalign{\smallskip} \toprule[0.8pt] \noalign{\smallskip}
\noalign{\smallskip}
$D_1D_1$ & \multirow{2}{*}{$0,0\oplus 1$}&$55\%$&\multirow{2}{*}{{$-3.0\pm0.1$}}  &\multirow{2}{*}{1.15}&\multirow{2}{*}{1.73}&\multirow{2}{*}{$-2.7$}&\multirow{2}{*}{$-5.2$}&\multirow{2}{*}{$0.1$}
&\multirow{2}{*}{$0.1$}&\multirow{2}{*}{$2.8$}&\multirow{2}{*}{$5.7$}&\multirow{2}{*}{$-3.8$}\\
\noalign{\smallskip}
$D^*_2D^*_2$ &                     &$45\%$&\\
\noalign{\smallskip}
\noalign{\smallskip}
$D_1D^*_2$ & \multirow{2}{*}{$0,1\oplus0$}&$56\%$&\multirow{2}{*}{{$-18.6\pm0.3$}}&\multirow{2}{*}{1.13}
&\multirow{2}{*}{1.34}&\multirow{2}{*}{$-4.1$}&\multirow{2}{*}{$-12.5$}&\multirow{2}{*}{$0.9$}&\multirow{2}{*}{$-8.5$}
&\multirow{2}{*}{$8.7$}&\multirow{2}{*}{$9.6$}&\multirow{2}{*}{$-12.7$}\\
\noalign{\smallskip}
$D^*_2D^*_2$ &                     &$44\%$&\\
\noalign{\smallskip}
\noalign{\smallskip}
$D_1D^*_2$ & \multirow{2}{*}{$0,1\oplus2$}&$71\%$&\multirow{2}{*}{{$-2.4\pm0.2$}}&\multirow{2}{*}{1.15}
&\multirow{2}{*}{2.02}&\multirow{2}{*}{$-3.0$}&\multirow{2}{*}{$-8.8$}&\multirow{2}{*}{$0.6$}&\multirow{2}{*}{$-0.8$}
&\multirow{2}{*}{$6.3$}&\multirow{2}{*}{$10.0$}&\multirow{2}{*}{$-6.7$}\\
\noalign{\smallskip}
$D^*_2D^*_2$ &                     &$29\%$&\\
\noalign{\smallskip}
\noalign{\smallskip}
$D_1D_1$ & \multirow{2}{*}{$1,0\oplus 0$}&$85\%$&\multirow{2}{*}{{$-16.4\pm0.2$}}  &\multirow{2}{*}{1.13}&\multirow{2}{*}{1.41}&\multirow{2}{*}{$-3.8$}&\multirow{2}{*}{$-1.3$}&\multirow{2}{*}{$-0.3$}
&\multirow{2}{*}{$-6.2$}&\multirow{2}{*}{$0.8$}&\multirow{2}{*}{$9.2$}&\multirow{2}{*}{$-14.9$}\\
\noalign{\smallskip}
$D^*_2D^*_2$ &                     &$15\%$&\\
\noalign{\smallskip}
\noalign{\smallskip}
$D_1D_1$ & \multirow{2}{*}{$1,0\oplus2$}&$95\%$&\multirow{2}{*}{{$-1.8\pm0.1$}}&\multirow{2}{*}{1.15}
&\multirow{2}{*}{2.25}&\multirow{2}{*}{$-2.5$}&\multirow{2}{*}{$-0.7$}&\multirow{2}{*}{$-0.2$}
&\multirow{2}{*}{$-0.0$}&\multirow{2}{*}{$-0.3$}&\multirow{2}{*}{$7.6$}&\multirow{2}{*}{$-5.7$}\\
\noalign{\smallskip}
$D^*_2D^*_2$ &                     &$5\%$&\\
\noalign{\smallskip}
\noalign{\smallskip}
$D_1D_2^*$   & $1,1\oplus 0$&Sole&{$-19.4\pm0.2$}&1.13&1.39&$-3.8$&$-1.3$&$-0.3$&$-8.7$&$-1.7$&$8.2$&$-11.8$\\
\noalign{\smallskip}
$D_1D_2^*$   & $1,1\oplus 2$&Sole&{$-3.6\pm0.1$} &1.15&1.99&$-3.0$&$-1.0$&$-0.2$&$-1.1$&$-1.4$&$9.3$&$-6.3$ \\
\noalign{\smallskip}
$D^*_2D^*_2$ & $1,2\oplus 0$&Sole&{$-19.6\pm0.2$}&1.14&1.40&$-3.8$&$-1.3$&$-0.3$&$-8.9$&$-1.9$&$8.2$&$-11.7$\\
\noalign{\smallskip}
$D^*_2D^*_2$ & $1,2\oplus 2$&Sole&{$-3.7\pm0.1$} &1.16&1.96&$-3.0$&$-1.0$&$-0.2$&$-1.2$&$-1.5$&$9.4$&$-6.3$ \\
\noalign{\smallskip}
\noalign{\smallskip}
\toprule[0.8pt]
\end{tabular}
\end{table*}

\subsection{S-wave $D_1D_1$ state}

It can be observed from Table~\ref{tcc} that the S-wave $D_1D_1$ state with the two isospin-spin-orbit
configurations $ISL=001$ ($IJ^P=01^+$) and $102$ ($IJ^P=12^+$) is not able to form a bound state
because of the absence of attraction within the quark model. In stark contrast, the S-wave $D_1D_1$
state, characterized by the configuration $ISL=100$, is predicted to form a bound state with a
binding energy of around 5.6 MeV relative to its constituent $D_1D_1$ in the quark model. In
this bound state, the majority of the attraction arises from the $\sigma$-meson exchange,
confinement potential and chromomagnetic interaction. Additionally, the kinetic energy, $\pi$-
and $\eta$-meson exchange provide small attractive contributions, less than 1 MeV. However,
the coulomb interaction introduces some repulsions, about 3.3 MeV.

Both the one-boson exchange model and chiral perturbation theory have also explored the
S-wave $D_1D_1$ state with $IJ^P=01^+$, $02^+$, and $10^+$ at the meson
level~\cite{Wang:2021ajy, Dong:2021bvy, Chen:2024snh}. Their conclusions contrast with the
present work, which finds that the isospin singlet is more likely to form a bound state
than the isospin triplet. The $D_1D_1$ state with $IJ^P=01^+$ can form a shallow bound
state with a molecular configuration when the cutoff parameter $\Lambda$ is
small~\cite{Wang:2021ajy, Dong:2021bvy, Chen:2024snh}. In Refs.~\cite{Dong:2021bvy, Wang:2021ajy},
the S-wave $D_1D_1$ state with $IJ^P=10^+$ and $12^+$ does not form a bound state. However,
in Ref.~\cite{Chen:2024snh}, while the S-wave $D_1D_1$ state with $IJ^P=10^+$ does not
establish a bound state, the S-wave $D_1D_1$ state with $IJ^P=12^+$ can form a loosely
bound state after including recoil corrections.

In general, the relative kinetic energy tends to be repulsive, preventing the constituent
particles from forming a bound state in few-body quantum
systems~\cite{Deng:2022vkv,Wei:2022wtr,Wang:2023vtx,Lin:2023gzm,Deng:2024pwm}.
The relative kinetic energy, acting as a novel binding mechanism in this bound state, can
be understood by examining the spatial configuration of the bound state. The size of the
two subclusters, $D_1D_1$ pair, is approximately 1.16 fm, and the distance between them
is about 1.89 fm. As a result, the two subclusters slightly overlap. This spatial overlap
increases the phase space available for the light quarks $\bar{q}_1$ and $\bar{q}_2$,
allowing them to move freely into the opposite subcluster. This contributes to the reduction
of the kinetic energy of the bound state, which exemplifies the realization of the
uncertainty principle.

The delocalization of the light quarks in the bound state closely resembles that of the electrons
in the hydrogen molecule, where the two hydrogen atoms are bound through the valence bond formed by
the delocalization of the two electrons. In this sense, the bound state $D_1D_1$ with $ISL=100$
is a hydrogenlike molecule state, where the QCD valence bond formed through the delocalization
of the light quarks although it is weaker than the $\sigma$-meson exchange, confinement
potential and chromomagnetic interaction. The formation of the QCD valence bond arises from the
exchange of light quarks, which ensures adherence to the Pauli exclusion principle. The concept
of the QCD valence bond was proposed by analogy with the hydrogen molecule in 2013~\cite{Du:2012wp}.
More recently, the QCD valence bond, along with other types of QCD bonds in multiquark hadrons,
has been extensively discussed in the literature~\cite{Chen:2021xlu,Chen:2025jgi}.

\subsection{S-wave $D_1D_2^*$ state}

The S-wave $D_1D_2^*$ state possesses six isospin-spin-orbit configurations, as illustrated in
Table~\ref{tcc}, due to the absence of constraints imposed by Bose-Einstein statistics. The
S-wave $D_1D_2^*$ state with $ISL=010$ ($IJ^P=01^+$), $011$ ($IJ^P=00^+$, $01^+$, $02^+$),
$012$ ($IJ^P=01^+$, $02^+$, $03^+$), and $111$ ($IJ^P=10^+$, $11^+$, $12^+$) cannot form
any bound states because of the lack of sufficient binding mechanisms within the quark model.

The S-wave $D_1D_2^*$ state with $ISL=110$ ($IJ^P=11^+$) can form a deeply bound state with
an approximate binding energy of 19.4 MeV within the quark model. In this bound state, the
majority of the attraction, about 11.8 MeV, comes from the kinetic energy, as shown in Table~\ref{tcc}.
Additionally, the confinement potential also provide a strong attraction, about 8.7 MeV.
The contributions from the meson exchanges ($\sigma$-, $\pi$-, and $\eta$-meson) and
chromomagnetic  interaction provide only small attractive effects. However, the Coulomb
interaction leads to a strongly repulsive effect of about 8.2 MeV.

The distance between the two subclusters, $D_1$ and $D_2^*$, is 1.39 fm, indicating a strong
overlap between the subclusters. By comparison, we observe that even if all other interactions
are switched off, leaving only the kinetic energy and Coulomb interaction, the system is still
able to form a bound state. In this sense, a strong QCD valence bond appears, making this bound
state resemble a hydrogen-like molecular state.

The S-wave $D_1 D_2^*$ state with $ISL=112$ ($IJ^P=11^+$, $12^+$, $13^+$) can generate a shallow
bound state, exhibiting an approximate binding energy of 3.6 MeV, in the quark model. The kinetic
energy still plays a decisive role in the formation of the bound state, as shown in Table~\ref{tcc}.
Additionally, the $\sigma$-meson exchange provides a strong attraction, contributing approximately
3.0 MeV. The meson exchanges ($\pi$ and $\eta$-meson), chromomagnetic interaction, and confinement
potential each contribute only small attractive effects. In contrast, the Coulomb interaction
results in a strongly repulsive effect, approximately 9.3 MeV. The distance between the two
subclusters, $D_1$ and $D_2^*$, is 1.99 fm, indicating a slightly overlap between the subclusters.
Consequently, this bound state can also be described as a hydrogen-like molecular state.

In the chiral perturbation theory, all isoscalar S-wave $D_1D_2^*$ states can form bound states,
whereas for all isovector states, the potential arising from $\rho$ and $\omega$ exchange is
repulsive, thereby preventing the formation of molecular states~\cite{Dong:2021bvy}. In the
one-boson-exchange model, the S-wave $D_1D_2^*$ state with $IJ^P=01^+$ and $02^+$ is easy to
form a bound state while the state with $13^+$ can generate a bound state under the condition
of large cutoff parameter~\cite{Wang:2021ajy}. These conclusions are in stark contrast to the
results obtained in the present study. The primary reason for these differences lies in the
obviously distinct binding mechanisms underlying the formation of the bound states.

\subsection{S-wave $D_2^*D_2^*$ state}

The S-wave $D_2^*D_2^*$ state has nine isospin-spin-orbit configurations, as shown in
Table~\ref{tcc}, because of the constraints imposed by Bose-Einstein statistics. Among these,
only three configurations $ISL=001$ ($IJ^P=01^+$), $120$ ($IJ^P=12^+$), and $122$
($IJ^P=10^+, 12^+, 14^+$) are capable of forming a bound state in the quark model. The
remaining configurations are unbound due to the absence of sufficient binding mechanisms.

The S-wave $D_2^*D_2^*$ state with $ISL=001$ is a shallow bound state, possessing a binding
energy of approximately 1.1 MeV. This state is notably distinct from the other bound states
discussed earlier, as the primary factor inhibiting the formation of the $D_2^*D_2^*$ pair
into a bound state is the kinetic energy. The distance between the two subclusters is
approximately 2.78 fm, which is greater than the size of each subcluster, suggesting that
this bound state resembles a deuteron-like molecule. In this bound state, the interaction
between the two subclusters arises from the exchange of $\pi$- and $\sigma$-mesons, along
with the confinement potential and chromomagnetic interaction. In the chiral perturbation
theory and one boson exchange model, the S-wave $D_2^*D_2^*$ state with $IJ^P=01^+$ and
$03^+$ can generate bound molecular states~\cite{Dong:2021bvy,Wang:2021ajy}.

The S-wave $D_2^*D_2^*$ state with $ISL=120$ can lead to the formation of a deep bound
state within the quark model, whereas the state with $ISL=122$ corresponds to a shallow
bound state. Upon comparison, we observe that their properties, including binding energy,
binding mechanisms, and spatial configurations, are strikingly similar to those of the
S-wave $D_1D_2^*$ state with $ISL=110$ and $ISL=112$, respectively, as shown in Table~\ref{tcc}.
As a result, the S-wave $D_2^*D_2^*$ states with $ISL=120$ and $122$ can be described as
hydrogen-like molecular states in the quark model. In the one-boson exchange model, the
S-wave $D_2^*D_2^*$ state with $IJ^P=14^+$ can form a bound molecular state~\cite{Wang:2021ajy}.
However, the states with $IJ^P=10^+$ and $12^+$ exhibit weak attraction or repulsion, depending
on the variation of the cutoff parameter within the range of 0.80 to 2.50 GeV~\cite{Wang:2021ajy}.

\subsection{Coupled channel effects}

From a quantum mechanical perspective, the dimeson states should be the linear combinations
of all possible isospin-spin-orbit configurations allowed by their quantum numbers. The
mixing of these configurations may lead to a lowering of the energy of the dimeson states.
In this context, we carry out coupling calculations involving the states that share the same
isospin-spin-orbit configurations in Table~\ref{tcc}. The numerical results of the coupling
calculations for the bound states are presented in Table~\ref{coupling} and Fig.~\ref{spectrum}.

The coupled binding energy for the S-wave $D_1D_1$ and $D_2^*D_2^*$ states with $ISL=001$
($IJ^P=01^+$) is approximately 3.0 MeV below the $D_1D_1$ threshold. In the bound state,
the contribution of $D_1D_1$ is slightly larger than that of $D_2^*D_2^*$. Similarly,
the S-wave $D_1D_2^*$ and $D_2^*D_2^*$ states with $ISL=010$ ($IJ^P=01^+$) and $012$
($IJ^P=01^+$, $03^+$) can also generate bound states relative to the $D_1D_2^*$ threshold
when the coupled-channel effect is taken into account. Their coupled binding energies are
18.6 MeV and 2.4 MeV, respectively, with the dominant component being $D_1D_2^*$. The
coupled-channel effect is essential in the formation of those bound states, significantly
influencing their existence.

The coupled binding energies for the S-wave $D_1D_1$ and $D_2^*D_2^*$ states with $ISL=100$
($IJ^P=10^+$) and $102$ ($IJ^P=20^+$) are approximately 16.4 MeV and 1.8 MeV lower than
the $D_1D_1$ threshold, respectively. In both bound states, the dominant component is $D_1D_1$,
as shown in Table~\ref{coupling}. The coupled-channel effect significantly lowers the
energy of the bound state $D_1D_1$ with $ISL=100$, even though it could form a bound
state independently. The formation of the shallow bound state with $ISL=102$ is entirely
dependent on the coupled-channel effect.

The exchange of $\pi$ and $\sigma$-meson plays a pivotal role in the bound states with
$I=0$, as they account for the majority of the attractive forces, as demonstrated in
Table~\ref{coupling}. Additionally, the kinetic energy contributes significantly to
the overall attraction. In contrast, the situation for the bound states with $I=1$
differs markedly. The kinetic energy serves as the primary binding mechanism in the
bound states with $I=1$, as it provides the majority of the attraction. In comparison,
the exchange of $\pi$ and $\sigma$-meson is relatively secondary even trivial. In this
context, it is challenging to form bound states with isospin $I=1$ at the meson
level, as the quark degrees of freedom are effectively frozen in both the one-boson
exchange model and chiral perturbation theory. In the bound states in Table~\ref{coupling},
the two subclusters overlap to some extent, making these bound states resemble
hydrogen-like molecules within the quark model.

\section{Summary}

In this study, we systematically investigate the properties of the S-wave states
$D_1D_1$, $D_1D^*_2$ and $D^*_2D^*_2$ with various isospin-spin-orbit configurations
in the quark model utilizing the Gaussian expansion method. By precisely solving the
fourbody Schr\"{o}dinger equations for the bound state question, we propose nine
possible bound states with the isospin-spin-orbit $(ISL)$ configurations, $ISL=001$,
$010$, $012$, $100$, $102$, $110$, $112$, $120$, and $122$, against dissociation
into their constituent mesons.

Those bound states are hydrogenlike molecular states, where the QCD covalent bond is
formed due to the delocalization of light quarks. The QCD covalent bond serves as the
primary binding mechanism in the bound states with $I=1$. Those bound states with $I=1$
cannot form in both the one-boson exchange model and chiral perturbation theory because
the quark degrees of freedom are effectively frozen. In contrast, the exchange of $\pi$
and $\sigma$-meson plays a pivotal role in the bound states with $I=0$. The
coupled-channel effect is essential in the formation of the bound states with $ISL=001$,
$010$, $012$, $100$, and $102$, significantly influencing their existence.

The information regarding the bound states presented in this work is expected to provide
valuable assistance and guidance for future experimental efforts aimed at exploring the
existence of exotic hadronic states.

\acknowledgments

This research is supported by the National Natural Science Foundation of China under
Grants No. 12305096, the Fundamental Research Funds for the Central Universities under Grant
No. SWU-XDJH202304 and No. SWU-KQ25016.

\end{document}